\documentclass[journal]{IEEEtran}
\usepackage{graphicx}
\usepackage{amsmath}
\usepackage{amssymb}
\usepackage{cite}
\usepackage{psfig}
\usepackage{subfigure}
\usepackage{booktabs}
\usepackage{pseudocode}
\renewcommand{\subfigcapskip}{0pt}
\renewcommand{\subfigbottomskip}{5pt}
\renewcommand{\subfigtopskip}{5pt}

\begin{document}

\title{Energy Efficient Transmission over Space Shift Keying Modulated MIMO Channels}

\author{Ronald Y. Chang, Sian-Jheng Lin, and Wei-Ho Chung,~\IEEEmembership{Member,~IEEE}%
\thanks{Paper approved by M. Juntti, the Editor for MIMO and Multiple-Access of the IEEE Communications Society. Manuscript received December 20, 2011; revised April 18, 2012.}%
\thanks{The authors are with the Research Center for Information Technology Innovation, Academia Sinica, Taipei, Taiwan (e-mail: \{rchang, sjlin, whc\}@citi.sinica.edu.tw).}%
\thanks{This work was supported by the National Science Council of Taiwan under Grant NSC 100-2221-E-001-004.}}

\markboth{IEEE Transactions on Communications, October 2012}%
{Chang \MakeLowercase{\textit{et al.}}: Energy Efficient Transmission over Space Shift Keying Modulated MIMO Channels}

\maketitle

\begin{abstract}
Energy-efficient communication using a class of spatial modulation (SM) that encodes the source information entirely in the antenna indices is considered in this paper. The energy-efficient modulation design is formulated as a convex optimization problem, where minimum achievable average symbol power consumption is derived with rate, performance, and hardware constraints. The theoretical result bounds any modulation scheme of this class, and encompasses the existing space shift keying (SSK), generalized SSK (GSSK), and Hamming code-aided SSK (HSSK) schemes as special cases. The theoretical optimum is achieved by the proposed practical energy-efficient HSSK (EE-HSSK) scheme that incorporates a novel use of the Hamming code and Huffman code techniques in the alphabet and bit-mapping designs. Experimental studies demonstrate that EE-HSSK significantly outperforms existing schemes in achieving near-optimal energy efficiency. An analytical exposition of key properties of the existing GSSK (including SSK) modulation that motivates a fundamental consideration for the proposed energy-efficient modulation design is also provided.
\end{abstract}


\section{Introduction} \label{sec:intro}

\IEEEPARstart{S}{patial} modulation (SM) is an emerging transmission technique that specifically exploits the deployment of multiple antennas in multiple-input multiple-output (MIMO) wireless communications. Unlike conventional phase and amplitude modulation such as quadrature amplitude modulation (QAM), SM encodes the source information partially or fully in the indices of the activated and idle transmit antennas. In the original SM scheme proposed in \cite{MeslehHaas08}, both the index of a single activated transmit antenna and the data symbol sent via the activated antenna carry information. The single activated antenna consideration is generalized to multiple activated antennas in \cite{YounisSerafimovski10,QuWang11,WangJia12,FuHou10} with sophisticated bit mapping rules. The limitation that the total number of transmit antennas has to be a power of two in the original SM scheme is relaxed in \cite{SerafimovskiDiRenzo10,YangAissa11} by proposing novel ways of fractional bit mapping. Trellis coded spatial modulation (TCSM) \cite{MeslehDiRenzo10,BasarAygolu11} improves the performance of SM in correlated channel conditions. Space-time block coded SM (STBC-SM) \cite{BasarAygolu11_2} combines space-time block coding (STBC) and SM. Analytical performance results for various SM schemes in different channel conditions are given in \cite{HandteMuller09,DiRenzoHaas10,DiRenzoHaas10_2,DiRenzoHaas11_2,DiRenzoHaas10_3}.

A class of SM that eliminates the use of phase and amplitude modulation and encodes the source information fully in the indices of transmit antennas has become a promising implementation of SM for future MIMO wireless communications. Space shift keying (SSK) \cite{JeganathanGhrayeb09} activates a single transmit antenna as in the original SM, which is generalized to a fixed number of multiple activated antennas in the generalized SSK (GSSK) \cite{JeganathanGhrayeb08}. Hamming code-aided SSK (HSSK) \cite{ChangLin11} links the constellation design of HSSK with the codeword construction technique of Hamming codes (in general, binary linear block codes) \cite{LinCostello04} and adopts a varying number of multiple activated antennas. Space-time shift keying (STSK) \cite{SugiuraChen10,SugiuraChen11} and space-time SSK (ST-SSK) \cite{YangXu11} extend SSK to both space and time dimensions by combining SSK with STBC. Opportunistic power allocation for SSK is suggested in \cite{DiRenzoHaas10_4} to improve the performance of SSK. SSK-type modulation presents an attractive class of modulation techniques, especially for future large-scale MIMO systems, for several reasons. First, the hardware cost in the radio frequency (RF) section \cite{MohammadiGhannouchi11} is reduced since only a subset of the transmit antennas are switched on for data transmission. The same hardware cost reduction benefits as in transmit antenna selection \cite{SanayeiNosratinia04} are achieved. Second, the detection complexity is lowered and the receiver design is simpler, since the information is contained entirely in the indexing of the antennas. Third, the transceiver requirement such as synchronization is reduced due to the absence of phase and amplitude modulation \cite{JeganathanGhrayeb09}. One disadvantage of SSK-type modulation is that the data rates increase only logarithmically with the number of transmit antennas, resulting in lower transmission rates as compared with conventional modulation. This problem can be alleviated by employing a large antenna array and activating multiple antennas, as well as by the efficient use of the set of antenna indices. An overview of the various aspects of SM (including SSK-type modulation) is available in \cite{DiRenzoHaas11}.

In this paper, we consider energy-efficient (green) communication using SSK-type modulation. Essentially, energy efficiency is achieved by nonequiprobable signaling where less power-consuming modulation symbols are used more frequently to transmit a given amount of information. We first derive the optimal transmission strategy that guarantees minimum average symbol power consumption provided that the spectral-efficiency, performance, and hardware constraints are met. The theoretical optimum is then achieved by the proposed energy-efficient HSSK (EE-HSSK) modulation scheme. The main contributions of this paper are summarized as follows.
\begin{itemize}
\item The dependency of the error performance of SSK-type modulation on the minimum Hamming distance between arbitrary two modulation symbols has been previously reported (see, e.g., \cite{ChangLin11}). In this paper, we show that the employment of a fixed number of activated antennas poses a fundamental limit on the performance of GSSK (including SSK) modulation. Specifically, we show that, among other properties of GSSK, the minimum Hamming distance between arbitrary two GSSK symbols is always equal to 2.
\item We investigate the important issue of energy-efficient transmission using SSK-type modulation. The proposed design framework considers a varied number of activated antennas, thus overcoming the limitation of GSSK modulation as mentioned above. The optimal transmission strategy is derived theoretically, which is achieved by the proposed practical EE-HSSK modulation scheme. The design of EE-HSSK is facilitated by a novel use of the bit-mapping technique (rather than the compression capability) of Huffman coding. The energy efficiency of EE-HSSK in comparison with existing schemes is verified by simulation. Various implementation issues of EE-HSSK are discussed.
\end{itemize}

This paper is organized as follows. Sec.~\ref{sec:system} presents the system description. The design problem and theoretical results for energy-efficient transmission are presented in Sec.~\ref{sec:method_theo}. The practical energy-efficient modulation scheme is described in Sec.~\ref{sec:method_prac}. Performance results are demonstrated in Sec.~\ref{sec:simulation}. Conclusion is given in Sec.~\ref{sec:conclusion}.

{\it Notations:} In this paper, $\mathbf I_N$ is the $N \times N$ identity matrix, $\left\| \cdot \right\|$ the $l_2$-norm of a vector, $(\cdot)^T$ and $(\cdot)^H$ the matrix/vector transpose and conjugate matrix/vector transpose, respectively, $Q(x)$ the Q-function defined as $\frac{1}{\sqrt{2\pi}}\int_x^{\infty}e^{-(\alpha^2/2)}d\alpha$, $|\cdot|$ the cardinality of a set, $\log$ the natural logarithm, $\log_2$ the base-2 logarithm, and $\lfloor x \rfloor$ the floor operation giving the largest integer not greater than $x$.

\section{Transmission Systems with SSK-Type Modulation} \label{sec:system}

\subsection{System Model} \label{sec:system1}

We consider an uncoded MIMO system with $N_T$ transmit antennas and $N_R$ receive antennas (denoted by an $N_T\times N_R$ system). The system employs an SSK-type modulation scheme that uses solely the antenna indices to carry information. The complex baseband signal model is given by
\begin{equation} \label{eq:model_c}
{\mathbf y} = {\mathbf H}\sqrt{E_s}\tilde{{\mathbf x}}+{\mathbf v}
\end{equation}
where ${\mathbf y}\in {\mathbb C}^{N_R\times 1}$ is the received signal, $\tilde{{\mathbf x}}$ is the $N_T\times 1$ transmitted symbol comprised of 1's (corresponding to activated antennas) and 0's (corresponding to idle antennas), ${\mathbf H}\in {\mathbb C}^{N_R\times N_T}$ is the flat-fading channel, ${\mathbf v}\in {\mathbb C}^{N_R\times 1}$ is the additive white Gaussian noise (AWGN), and $E_s$ is the transmit power at each antenna. Channel matrix ${\mathbf H}$ has independent and identically distributed (i.i.d.) complex Gaussian entries with zero mean and covariance matrix $\sigma_H^2 {\mathbf I}_{N_R}$, where $\sigma_H^2=1$. The channel information is assumed perfectly known to the receiver. Noise ${\mathbf v}$ has i.i.d. complex elements with zero mean and covariance matrix $(N_0/2) {\mathbf I}_{N_R}$. Transmitted symbol $\tilde{{\mathbf x}}$ is selected from the modulation alphabet (constellation set) ${\cal A}$ with {\it a priori} probability $P(\tilde{{\mathbf x}})$, where $\sum_{\tilde{{\mathbf x}}\in {\cal A}} P(\tilde{{\mathbf x}})=1$. 

\subsection{System Performance} \label{sec:system2}

Given the signal model in (\ref{eq:model_c}), optimal detection is the maximum {\it a posteriori} probability (MAP) estimate, i.e.,
\begin{equation}
\tilde{{\mathbf x}}_{\mbox{\scriptsize MAP}}=\arg\max_{{\mathbf x}\in {\cal A}} p({\mathbf y}|{\mathbf x})P({\mathbf x})
\end{equation}
or equivalently
\begin{equation}
\tilde{{\mathbf x}}_{\mbox{\scriptsize MAP}}=\arg\max_{{\mathbf x}\in {\cal A}} \log\Big(p({\mathbf y}|{\mathbf x})P({\mathbf x})\Big)
\end{equation}
where
\begin{equation}
p({\mathbf y}|{\mathbf x})=\frac{1}{\pi^{N_R}\big(\frac{N_0}{2}\big)^{N_R}} \exp\Big(-\frac{2}{N_0}\left\|{\mathbf y} - {\mathbf H}\sqrt{E_s}{\mathbf x}\right\|^2\Big).
\end{equation}
Therefore,
\begin{equation} \label{eq:MAP}
\tilde{{\mathbf x}}_{\mbox{\scriptsize MAP}}=\arg\min_{{\mathbf x}\in {\cal A}} \Big(\left\|{\mathbf y} - {\mathbf H}\sqrt{E_s}{\mathbf x}\right\|^2-\log P({\mathbf x})\Big).
\end{equation}
When alphabet ${\cal A}$ contains equiprobable elements, i.e., $P({\mathbf x})=1/|{\cal A}|$ for all ${\mathbf x}$, MAP detection reduces to maximum likelihood (ML) detection which does not incorporate the {\it a priori} probabilities in its detection metric \cite{JeganathanGhrayeb08_2}. 

The system performance based on MAP detection can be quantified by first deriving the pairwise error probability (PEP) and then averaging the PEP over all pairwise symbol combinations. Let ${\mathbf x}_i$ and ${\mathbf x}_j$ be two distinct symbols in ${\cal A}$. Following the same derivation approach in \cite{JeganathanGhrayeb08,ChangLin11}, the PEP of deciding on ${\mathbf x}_j$ given that ${\mathbf x}_i=\tilde{{\mathbf x}}$ is transmitted is given by
\begin{equation}
P({\mathbf x}_i\rightarrow {\mathbf x}_j | {\mathbf x}_i=\tilde{{\mathbf x}}) = \int_0^{\infty} Q\Big(\sqrt{x}+\frac{L({\mathbf x}_i, {\mathbf x}_j)}{N_0\sqrt{x}}\Big) f_Z(x)\,dx \label{eq:PEP1}
\end{equation}
where $L({\mathbf x}_i, {\mathbf x}_j)\triangleq \log\big(P({\mathbf x}_i)/P({\mathbf x}_j)\big)$ is the log-ratio between the {\it a priori} probability of ${\mathbf x}_i$ and ${\mathbf x}_j$, $f_Z$ is the probability density function (pdf) of $Z=\sum_{k=1}^{2N_R} z_k^2$, where $z_k$'s are i.i.d. ${\cal N}(0,\sigma_z^2)$ with $\sigma_z^2=E_sd({\mathbf x}_i, {\mathbf x}_j)/(2N_T N_0)$, and $d({\mathbf x}_i, {\mathbf x}_j)$ is the Hamming distance between ${\mathbf x}_i$ and ${\mathbf x}_j$. The system error probability $P_s$ can be derived by averaging the result in (\ref{eq:PEP1}) over all pairwise combinations, i.e.,
\begin{equation} \label{eq:SER}
P_s = \sum_{{\mathbf x}_i\neq {\mathbf x}_j\in{\cal A}} P({\mathbf x}_i)P({\mathbf x}_j|{\mathbf x}_j\neq {\mathbf x}_i)P({\mathbf x}_i\rightarrow {\mathbf x}_j | {\mathbf x}_i=\tilde{{\mathbf x}}). 
\end{equation}
The dominant terms in (\ref{eq:SER}) involve the error events of symbol ${\mathbf x}_i$ with the largest {\it a priori} probability $P({\mathbf x}_i)$. The PEP in these dominant terms can be obtained by substituting the known expression of $f_Z$ into (\ref{eq:PEP1}) and using the Chernoff bound $Q(x)\leq (1/2)\exp(-x^2/2),x\geq 0$ in (\ref{eq:PEP1}) (since $L({\mathbf x}_i, {\mathbf x}_j)\geq 0$), i.e.,
\begin{eqnarray}
&& \hspace{-0.34in} P({\mathbf x}_i\rightarrow {\mathbf x}_j | {\mathbf x}_i=\tilde{{\mathbf x}}) \nonumber\\
&& \hspace{-0.3in} \leq \int_0^{\infty} \frac{1}{2}\exp\Big(-\frac{x}{2}\Big) \exp\Big(-\frac{L({\mathbf x}_i, {\mathbf x}_j)}{N_0}\Big) \nonumber\\
&& \hspace{1.2in} \cdot\exp\Big(-\frac{L({\mathbf x}_i, {\mathbf x}_j)^2}{2N_0^2 x}\Big) f_Z(x)\,dx \nonumber\\
&& \hspace{-0.3in} \leq \frac{1}{2}\exp\Big(-\frac{L({\mathbf x}_i, {\mathbf x}_j)}{N_0}\Big) \int_0^{\infty} \exp\Big(-\frac{x}{2}\Big) f_Z(x)\,dx \nonumber\\
&& \hspace{-0.3in} = \frac{1}{2}\exp\Big(-\frac{L({\mathbf x}_i, {\mathbf x}_j)}{N_0}\Big)\cdot \big(\sigma_z^2+1\big)^{-N_R} \nonumber\\
&& \hspace{-0.3in} = \frac{1}{2}\exp\Big(-\frac{L({\mathbf x}_i, {\mathbf x}_j)}{N_0}\Big)\cdot \Big(\frac{E_s}{N_0}\cdot d({\mathbf x}_i, {\mathbf x}_j)\cdot \frac{1}{2N_T}+1\Big)^{-N_R} \nonumber\\ \label{eq:PEP2}
\end{eqnarray}
where for the second inequality we have used the fact that $\exp\big(-L({\mathbf x}_i, {\mathbf x}_j)^2\!/(2N_0^2 x)\big)\leq 1$ in the domain of integration. Note that the bound in (\ref{eq:PEP2}) has been shown for equiprobable signaling \cite{ChangLin11} to be loose in low signal-to-noise ratio (SNR) and tighter in high SNR regimes. Substituting (\ref{eq:PEP2}) in (\ref{eq:SER}) it is clear that at a given operating $E_s/N_0$ the system performance is dominated by pairs of symbols with large {\it a priori} probabilities and with small pairwise Hamming distances. When the {\it a priori} probabilities are all equal, the minimum Hamming distance between arbitrary two distinct symbols, i.e., $d_{\mbox{\scriptsize min}}=\min_{{\mathbf x}_i\neq {\mathbf x}_j\in{\cal A}}d({\mathbf x}_i, {\mathbf x}_j)$, determines the system performance. When the {\it a priori} probabilities are not all equal where symbols with large (small) {\it a priori} probabilities have smaller (larger) pairwise Hamming distances, again, the minimum Hamming distance between arbitrary two distinct symbols determines the system performance. In our energy-efficient consideration for SSK-type modulation enabled transmission (Sec.~\ref{sec:method_theo}), symbols with large {\it a priori} probabilities typically have smaller pairwise Hamming distances, since creating large Hamming distances requires the employment of large numbers of 1's in the resultant more power-consuming modulation symbols. Note that the analytical insights established here apply to the uncoded system. For the performance analysis for coded systems, the reader is referred to \cite{MeslehDiRenzo10,JeganathanGhrayeb09}. Specifically, the convolutional coded bit error probability is shown to be bounded by a function of PEP in \cite[Sec. IV-B]{MeslehDiRenzo10} and \cite[Sec. IV-D]{JeganathanGhrayeb09}, with the PEP analyzed through the Hamming distance.

\section{Energy-Efficient HSSK Transmission: Theoretical Results} \label{sec:method_theo}

\subsection{Motivation of the Proposed Design}

A well-known SSK-type modulation is GSSK modulation \cite{JeganathanGhrayeb08}. To support $m$ bits/s/Hz transmission in a system with $N_T$ transmit antennas, GSSK modulation employs a fixed number of $n_t$ 1's and $N_T-n_t$ 0's in each $N_T\times 1$ modulation symbol, where $n_t$ is chosen such that ${N_T \choose n_t-1} < |{\cal A}^{(\mbox{\scriptsize GSSK})}|=2^m \leq {N_T \choose n_t}$ and $n_t\leq N_T/2$. The employment of $n_t$ being the minimum number of 1's required to encode the given amount of information ensures energy efficiency and conservation of the hardware expenses (i.e., the required number of RF chains at the transmitter). An example of the GSSK alphabet is shown in Table~\ref{tab:mod} for a system with $N_T=5$ and $n_t=2$ to support $m=3$ bits/s/Hz transmission. If $|{\cal A}^{(\mbox{\scriptsize GSSK})}|<{N_T \choose n_t}$ as in this example, ${\cal A}^{(\mbox{\scriptsize GSSK})}$ is not unique and may be chosen lexicographically as shown in Table~\ref{tab:mod} or based on some optimization criterion at an extra computational cost \cite{JeganathanGhrayeb08}.

%%%%%%%%%%%%%%%%%%%%%%%%%%%%%%%%%%%%%%%%%%%%%%%%%%%%%%%%%%%%%%%%%
\begin{table}[tb]
\begin{center}
\caption{Example of GSSK ($n_t=2$) Modulation Alphabet and Bit Mapping for 3 Bits/s/Hz Transmission in a System with $N_T=5$}
\label{tab:mod} \vspace*{1mm}
\begin{tabular}{|l|c|}
\hline
Source bits & GSSK symbols $\in {\cal A}^{(\mbox{\tiny GSSK})}$ \\\hline\hline
$0 0 0$ & $[0,0,0,1,1]^T$ \\\hline
$0 0 1$ & $[0,0,1,0,1]^T$ \\\hline
$0 1 0$ & $[0,1,0,0,1]^T$ \\\hline
$0 1 1$ & $[1,0,0,0,1]^T$ \\\hline
$1 0 0$ & $[0,0,1,1,0]^T$ \\\hline
$1 0 1$ & $[0,1,0,1,0]^T$ \\\hline
$1 1 0$ & $[1,0,0,1,0]^T$ \\\hline
$1 1 1$ & $[0,1,1,0,0]^T$ \\\hline
\end{tabular}
\end{center}
\end{table}
%%%%%%%%%%%%%%%%%%%%%%%%%%%%%%%%%%%%%%%%%%%%%%%%%%%%%%%%%%%%%%%%%

While useful for improving the transmission rates of SSK modulation (a special case of GSSK with $n_t=1$), GSSK modulation presents several limitations.
\begin{itemize}
\item[1)] {\it Transmission rate:} The employment of a fixed number of $n_t$ 1's in each modulation symbol results in an inefficient utilization of the set of antenna indices and poses limitation on the attainable transmission rates of GSSK.
\item[2)] {\it Symbol selection:} When the size of the GSSK modulation alphabet $|{\cal A}^{(\mbox{\scriptsize GSSK})}|=2^m$ is smaller than the number of available symbols ${N_T \choose n_t}$, a sophisticated selection of symbols into ${\cal A}^{(\mbox{\scriptsize GSSK})}$ (as compared with plain lexicographic selection) typically requires a computationally intense search over the ${{N_T\choose n_t} \choose 2^m}$ possibilities of the alphabet.
\item[3)] {\it System performance:} The GSSK modulation alphabet, constructed based on lexicographic selection or any other criterion, always leads to $d_{\mbox{\scriptsize min}}=\min_{{\mathbf x}_i\neq {\mathbf x}_j\in{\cal A}^{(\mbox{\scriptsize GSSK})}}d({\mathbf x}_i, {\mathbf x}_j)=2$. This limits the performance of GSSK modulation as discussed previously in Sec.~\ref{sec:system2}. Furthermore, an increasingly small subset of the total ${N_T \choose n_t}$ available symbols can have a pairwise Hamming distance larger than 2 as $N_T$ increases. These performance-related properties of GSSK modulation are analytically explored in the Appendix.
\end{itemize}

The perceived limitations of GSSK modulation can be overcome by fully utilizing the set of antenna indices, as implemented by HSSK modulation \cite{ChangLin11}. HSSK modulation employs a varied number of 1's in each modulation symbol based on the Hamming code (in general, binary linear block code) construction technique. For example, to support 4 bits/s/Hz transmission in a system with $N_T=5$, the HSSK modulation alphabet incorporates the 16 codewords of the $(5,4)$ code\footnote{We use the notation $(n,k)$ for binary linear block codes with block length $n$, message length $k$, and an implicit minimum distance $d$. For example, $(n,n-1)$ is binary parity check code with $d=2$, and $(n,1)$ is repetition code with $d=n$.} with the last bit of each codeword complemented, as shown in Table \ref{tab:mod2}. Note that 4 bits/s/Hz transmission is achievable with HSSK but not with GSSK in a system with $N_T=5$. Furthermore, as opposed to GSSK with $d_{\mbox{\scriptsize min}}=2$ always, HSSK can be configured with a flexible $d_{\mbox{\scriptsize min}}$. For example, the HSSK modulation alphabet may incorporate the four codewords of the $(5,2)$ code $[0,0,0,0,1]^T$, $[0,0,1,1,0]^T$, $[1,1,0,0,0]^T$, and $[1,1,1,1,1]^T$ to achieve $d_{\mbox{\scriptsize min}}=3$ in support of 2 bits/s/Hz transmission in a system with $N_T=5$. Clearly, there is a tradeoff in terms of achievable rate, $d_{\mbox{\scriptsize min}}$, power and hardware costs in the selection of design options. For a detailed coverage on the comparison of HSSK and GSSK, the reader is referred to \cite{ChangLin11}.

Since HSSK modulation employs a varied number of 1's in the modulation symbols to fully utilize the set of antenna indices, some modulation symbols naturally require higher transmission power and more activated transmit antennas than others. This motivates an energy-efficient and hardware-cost-aware refinement of HSSK-enabled transmission that maintains the advantages of HSSK modulation. The design problem and theoretical results are presented next, and a practical energy-efficient HSSK transmission scheme is presented in Sec.~\ref{sec:method_prac}.

%%%%%%%%%%%%%%%%%%%%%%%%%%%%%%%%%%%%%%%%%%%%%%%%%%%%%%%%%%%%%%%%%
\begin{table}[tb]
\begin{center}
\caption{Example of HSSK Modulation Alphabet and Bit Mapping for 4 Bits/s/Hz Transmission in a System with $N_T=5$}
\label{tab:mod2} \vspace*{1mm}
\begin{tabular}{|l|c|}
\hline
Source bits & HSSK symbols $\in {\cal A}^{(\mbox{\tiny HSSK})}$ \\\hline\hline
$0 0 0 0$ & $[0,0,0,0,1]^T$ \\\hline
$0 0 0 1$ & $[0,0,0,1,0]^T$ \\\hline
$0 0 1 0$ & $[0,0,1,0,0]^T$ \\\hline
$0 0 1 1$ & $[0,1,0,0,0]^T$ \\\hline
$0 1 0 0$ & $[1,0,0,0,0]^T$ \\\hline
$0 1 0 1$ & $[0,0,1,1,1]^T$ \\\hline
$0 1 1 0$ & $[0,1,0,1,1]^T$ \\\hline
$0 1 1 1$ & $[1,0,0,1,1]^T$ \\\hline
$1 0 0 0$ & $[0,1,1,0,1]^T$ \\\hline
$1 0 0 1$ & $[1,0,1,0,1]^T$ \\\hline
$1 0 1 0$ & $[1,1,0,0,1]^T$ \\\hline
$1 0 1 1$ & $[0,1,1,1,0]^T$ \\\hline
$1 1 0 0$ & $[1,0,1,1,0]^T$ \\\hline
$1 1 0 1$ & $[1,1,0,1,0]^T$ \\\hline
$1 1 1 0$ & $[1,1,1,0,0]^T$ \\\hline
$1 1 1 1$ & $[1,1,1,1,1]^T$ \\\hline
\end{tabular}
\end{center}
\end{table}
%%%%%%%%%%%%%%%%%%%%%%%%%%%%%%%%%%%%%%%%%%%%%%%%%%%%%%%%%%%%%%%%%

\subsection{Problem Formulation}

Let ${\cal U}^s$ be the $N_T$-dimensional binary symbol universe, i.e., ${\cal U}^s=\big\{[x_1,x_2,\ldots,x_{N_T}]^T|x_j\in\{0,1\}, j=1,\ldots,N_T\big\}$, and ${\cal U}_i$ ($i=0,1,\ldots,N_T$) be a subset of ${\cal U}^s$ containing symbols with exactly $i$ 1's, i.e., ${\cal U}_i=\big\{[x_1,x_2,\ldots,x_{N_T}]^T|x_j\in\{0,1\}, j=1,\ldots,N_T, \sum_{j=1}^{N_T} x_j=i\big\}$. The HSSK modulation alphabet with a specified $d_{\mbox{\scriptsize min}}$ property incorporates selected codewords in the code ${\cal C}=\bigcup_i {\cal C}_i$ with minimum distance $d=d_{\mbox{\scriptsize min}}$, where ${\cal C}_i\subseteq {\cal U}_i$ ($i\in\{1,\ldots,N_T\}$) is the nonempty set of codewords containing exactly $i$ 1's. The all-zero symbol in ${\cal U}_0$ is not used in the modulation alphabet and is reserved for special use (see Sec.~\ref{sec:method_prac2}). For example, in the previously described example, the HSSK modulation alphabet incorporates all codewords in the $(5,2)$ code ${\cal C}={\cal C}_1\bigcup{\cal C}_2\bigcup{\cal C}_5$ with $d=d_{\mbox{\scriptsize min}}=3$ to achieve 2 bits/s/Hz transmission in a system with $N_T=5$, where $|{\cal C}_1|=|{\cal C}_5|=1$ and $|{\cal C}_2|=2$. It is often useful to further restrict the required number of RF chains at the transmitter to $M$ ($1\leq M\leq N_T$) for the modulation scheme. Therefore, the objective of energy-efficient transmission based on HSSK modulation is to design an alphabet and the symbol {\it a priori} probabilities so that minimum average symbol power per transmission ${\mathbb E}[\tilde{{\mathbf x}}^H \tilde{{\mathbf x}}]$ is achieved, while the target transmission rate (spectral-efficiency constraint), the minimum Hamming distance property (performance constraint), and the maximum required number of RF chains (hardware constraint) are met. Given a code ${\cal C}=\bigcup_i {\cal C}_i$ with the specified minimum distance property, and given the fact that each element in ${\cal C}_i$ requires $i$ RF chains at the transmitter and consumes power equal to $i$, the design problem is mathematically formulated as
\begin{eqnarray}
\min_{P_i} && \hspace{-0.1in} \sum_{\substack{i: {\cal C}_i\subseteq {\cal C} \\ i\leq M}} i|{\cal C}_i|P_i \nonumber\\
\mbox{s.t.} && \hspace{-0.1in} \sum_{\substack{i: {\cal C}_i\subseteq {\cal C} \\ i\leq M}} |{\cal C}_i|P_i=1 \nonumber\\
&& \hspace{-0.1in} \sum_{\substack{i: {\cal C}_i\subseteq {\cal C} \\ i\leq M}} |{\cal C}_i|r(P_i)\geq m \label{eq:opt}
\end{eqnarray}
where $P_i$ is the {\it a priori} probability of each symbol in ${\cal C}_i$, i.e., $P_i\triangleq P({\mathbf x}), {\mathbf x}\in {\cal C}_i$, and
\begin{equation}
r(P_i)=\left\{\begin{matrix}
-P_i\log_2 P_i, & \mbox{if } P_i>0 \\ 
0, & \mbox{if } P_i=0 \\ 
-\infty, & \mbox{otherwise}
\end{matrix}\right..
\end{equation}
The first constraint of Problem (\ref{eq:opt}) states that the {\it a priori} probabilities of all symbols in the alphabet sum to one, and implies that $P_i=0$ for $i$ for which ${\cal C}_i\nsubseteq {\cal C}$ or $i>M$. The second constraint states that the target information rate of $m$ bits is met, described by Shannon's entropy formula. Clearly, the optimal power yielded in Problem (\ref{eq:opt}) with a larger $M$ is no greater than with a smaller $M$.

Consider an example. Suppose that it is desired to design a modulation alphabet with $d_{\mbox{\scriptsize min}}=2$. The largest set of length-$N_T$ codewords with $d=d_{\mbox{\scriptsize min}}=2$ is the $(N_T,N_T-1)$ binary parity check code. More specifically, the code ${\cal C}$ is given by ${\cal C}_1\bigcup{\cal C}_3\bigcup{\cal C}_5\bigcup\cdots\bigcup{\cal C}_{N_T}$ if $N_T$ is odd and ${\cal C}_1\bigcup{\cal C}_3\bigcup{\cal C}_5\bigcup\cdots\bigcup{\cal C}_{N_T-1}$ if $N_T$ is even, where ${\cal C}_i={\cal U}_i$. Since $|{\cal C}_i|=|{\cal U}_i|={N_T\choose i}$, the design problem is specifically formulated as
\begin{eqnarray}
\min_{\{P_1, P_3, P_5, \ldots\}} && \hspace{-0.1in} \sum_{i=1, 3, 5,\ldots}^M i{N_T\choose i}P_i \nonumber\\
\mbox{s.t.} && \hspace{-0.1in} \sum_{i=1, 3, 5,\ldots}^M {N_T\choose i}P_i=1 \nonumber\\
&& \hspace{-0.1in} \sum_{i=1, 3, 5,\ldots}^M {N_T\choose i}r(P_i)\geq m. \label{eq:opt2}
\end{eqnarray}
The first constraint of Problem (\ref{eq:opt2}) implies that $P_2,P_4,\ldots$ are all equal to zero, meaning that symbols in ${\cal U}_2,{\cal U}_4,\ldots$ are not included in the modulation alphabet.

\subsection{Optimal Solution}

Note that Problem (\ref{eq:opt}) has a linear objective subject to affine equality and convex inequality constraints. Therefore, they are convex optimization problems with a globally optimal solution, and can be solved using the Lagrange multiplier method \cite{Bertsekas99}. The Lagrangian function for Problem (\ref{eq:opt}) is given by
\begin{eqnarray}
\Lambda(P_i,\lambda_1,\lambda_2)=\sum_{\substack{i: {\cal C}_i\subseteq {\cal C} \\ i\leq M}} i|{\cal C}_i|P_i+\lambda_1\Bigg(\sum_{\substack{i: {\cal C}_i\subseteq {\cal C} \\ i\leq M}} |{\cal C}_i|P_i-1\Bigg) \nonumber\\
+\lambda_2\Bigg(\sum_{\substack{i: {\cal C}_i\subseteq {\cal C} \\ i\leq M}} |{\cal C}_i|P_i\log_2 P_i+m\Bigg)
\end{eqnarray}
where $\lambda_1$ and $\lambda_2$ are the Lagrange multipliers. The optimal solution $P_i^*$ and the associated Lagrange multipliers $\lambda_1^*$ and $\lambda_2^*$ satisfy the following Karush-Kuhn-Tucker (KKT) necessary conditions
\begin{equation} \label{eq:lagrange}
\left\{\begin{array}{l}
\partial\Lambda(P_i^*,\lambda_1^*,\lambda_2^*)/\partial P_i=0, \quad i:{\cal C}_i\subseteq {\cal C},i\leq M \\
\lambda_2^*\geq 0 \\
\lambda_2^*=0,\quad \mbox{if the inequality constraint of (\ref{eq:opt}) is inactive}.
\end{array}\right.
\end{equation}
By directly computing the first condition in (\ref{eq:lagrange}) we have
\begin{equation} \label{eq:lagrange2}
i|{\cal C}_i|+\lambda_1^*|{\cal C}_i|+\lambda_2^*|{\cal C}_i|\Big(\log_2 P_i^*+\frac{1}{\log 2}\Big)=0.
\end{equation}
Arranging the terms yields $P_i^*=\alpha \beta^i$, where $\alpha=(1/e) 2^{-\lambda_1^*/\lambda_2^*}$ and $\beta=2^{-1/\lambda_2^*}$. Normalizing $P_i^*$'s according to the equality constraint of Problem (\ref{eq:opt}), we obtain
\begin{equation} \label{eq:lagrange3}
P_i^*=\frac{\beta^i}{\sum_{i: {\cal C}_i\subseteq {\cal C},i\leq M} |{\cal C}_i|\beta^i}, \quad 0<\beta\leq 1.
\end{equation}
The value of $\beta$ determines the optimal {\it a priori} probabilities for the alphabet. If $\beta=1$, we have $P_i^*=1/\big(\sum_{i: {\cal C}_i\subseteq {\cal C},i\leq M} |{\cal C}_i|\big)$ for all $i$ for which ${\cal C}_i\subseteq {\cal C}$ and $i\leq M$, i.e., all codewords in ${\cal C}$ are included in the alphabet equiprobably to achieve the highest information rate of $\log_2\big(\sum_{i: {\cal C}_i\subseteq {\cal C},i\leq M} |{\cal C}_i|\big)$ bits at the cost of the largest average symbol power consumption. If $\beta=0^+$, we have $P_i^*=1/|{\cal C}_i|$ for the smallest $i$ for which ${\cal C}_i\subseteq {\cal C}$ and $P_i^*=0$ otherwise, i.e., only the least power-consuming codewords in ${\cal C}$ are included in the alphabet equiprobably to support an information rate as high as $\log_2 |{\cal C}_i|$ bits, where $i$ is the smallest index for which ${\cal C}_i\subseteq {\cal C}$. If the target information rate $m$ is less than this rate, the inequality constraint of Problem (\ref{eq:opt}) becomes inactive at the same optimal solution point.

The solution to Problem (\ref{eq:opt2}) is given similarly:
\begin{equation} \label{eq:lagrange4}
P_i^* = \frac{\beta^i}{\sum_{i=1, 3, 5,\ldots}^M {N_T\choose i}\beta^i}, \quad 0<\beta\leq 1, i=1,3,5,\ldots.
\end{equation}
Note that the solution in (\ref{eq:lagrange4}) coincides with the original GSSK and HSSK ($d_{\mbox{\scriptsize min}}=2$) modulation in scenarios in which the original GSSK or HSSK is energy-optimal. For example, to support 4 bits/s/Hz transmission in a system with $N_T=5$, the solution to Problem (\ref{eq:opt2}) with $M=N_T=5$ and $m=4$ is given by $P_1^*=P_3^*=P_5^*=1/16$ in (\ref{eq:lagrange4}) with $\beta=1$. This coincides with the original HSSK alphabet shown in Table \ref{tab:mod2} incorporating symbols in ${\cal U}_1$, ${\cal U}_3$, and ${\cal U}_5$ equiprobably. For another example, to support 2 bits/s/Hz transmission in a system with $N_T=4$, the solution to Problem (\ref{eq:opt2}) with $M=N_T=4$ and $m=2$ is given by $P_1^*=1/4$ and $P_3^*=0$ in (\ref{eq:lagrange4}) with $\beta=0^+$, which coincides with the SSK alphabet incorporating symbols in ${\cal U}_1$ equiprobably.

Problems (\ref{eq:opt}) and (\ref{eq:opt2}) can alternatively be formulated so that the objective is to maximize the transmission rate under the average symbol power constraint (with other constraints unchanged). It is not difficult to see that the same solution set as (\ref{eq:lagrange3})--(\ref{eq:lagrange4}) will be yielded. In other words, the proposed design scheme will either achieve higher transmission rate at the same symbol power consumption, or achieve less symbol power consumption at the same target transmission rate, than any other SSK-type modulation scheme. The theoretical study in this section provides the optimal symbol {\it a priori} probabilities. However, no information is given for accomplishing the bit mapping as in Tables \ref{tab:mod}--\ref{tab:mod2}. In practice, the technique of variable-length coding is useful for creating an efficient bit-string representation of symbols with unequal {\it a priori} probabilities, as presented next.

\section{Energy-Efficient HSSK Transmission: Practical Schemes} \label{sec:method_prac}

In this section, we present the proposed practical EE-HSSK transmission scheme and discuss its implementation issues.

\subsection{EE-HSSK Transmission Model}

Huffman coding \cite{Huffman52} is an entropy coding algorithm for encoding a source of symbols with unequal symbol frequencies. The algorithm achieves a mapping from symbols to bit strings that has the smallest average bit-string length among all binary codes. Therefore, it has been widely used for data compression. The algorithm first generates a symbol frequency-sorted binary tree (i.e., the Huffman tree), and then assigns each symbol with a unique bit string based on the structure of the tree. Here, instead of using the compression capability of Huffman coding, we apply its bit-mapping technique in the proposed modulation scheme.

\begin{figure}[tb]
\begin{center}
\includegraphics[width=\columnwidth]{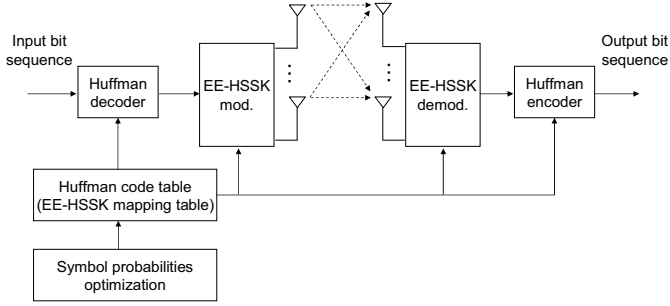}
\caption{System model for EE-HSSK modulation enabled transmission.} \label{fig:HSSKmodel}
\end{center}
\end{figure}

The system model for EE-HSSK enabled transmission is shown in Fig.~\ref{fig:HSSKmodel}. At the transmitter, Problem (\ref{eq:opt}) is first solved to obtain the optimal symbol probabilities. Then, the Huffman tree is built for these symbol probabilities, and the Huffman code table (used as the modulation mapping table) is constructed. The Huffman decoder then maps the input bit sequence to the modulation symbols according to the mapping table, and the EE-HSSK modulator and the RF front-end transmit the signal using the corresponding modulation symbols. At the receiver, the EE-HSSK demodulator detects the signal using the MAP criterion, and the Huffman encoder maps each detected symbol to bits according to the mapping table to generate the output bit sequence.

%%%%%%%%%%%%%%%%%%%%%%%%%%%%%%%%%%%%%%%%%%%%%%%%%%%%%%%%%%%%%%%%%
%%%%%%%%%%%%%%%%%%%%%%%%%%%%%%%%%%%%%%%%%%%%%%%%%%%%%%%%%%%%%%%%%

\subsection{Implementation Details} \label{sec:method_prac2}

An example of EE-HSSK modulation alphabet is shown in Table \ref{tab:mod3} for (approximately) 3 bits/s/Hz transmission in a system with $N_T=5$. This alphabet is given by first solving Problem (\ref{eq:opt2}) with $M=3$ to obtain the optimal symbol probabilities and then using Huffman coding to approximate the symbol probabilities. As can be clearly seen, unlike standard ASCII coding-based bit mapping in Tables \ref{tab:mod}--\ref{tab:mod2}, a Huffman code-aided bit mapping is employed here where the length of the bit strings is roughly reversely proportional to the symbol power. Since longer bit strings appear less frequently in a random input sequence, symbols more power-consuming are used less frequently to achieve energy efficiency. Note that the optimal {\it a priori} probability given by Problem (\ref{eq:opt2}) for a symbol is only approximately matched by probability $2^{-k}$ using Huffman coding, where $k$ is the length of the bit string associated with the symbol. Therefore, EE-HSSK modulation may achieve a slightly different transmission rate than is configured in Problem (\ref{eq:opt}) or (\ref{eq:opt2}), and may incur an energy penalty compared to the optimal theoretical results. The energy penalty, however, is generally very small (see Sec.~\ref{sec:simulation}).

%%%%%%%%%%%%%%%%%%%%%%%%%%%%%%%%%%%%%%%%%%%%%%%%%%%%%%%%%%%%%%%%%
\begin{table}[tb]
\begin{center}
\caption{Example of EE-HSSK ($M=3$) Modulation Alphabet and Bit Mapping for Approximately 3 Bits/s/Hz Transmission in a System with $N_T=5$}
\label{tab:mod3} \vspace*{1mm}
\begin{tabular}{|l|c|}
\hline
Source bits & EE-HSSK symbols $\in {\cal A}^{(\mbox{\tiny EE-HSSK})}$ \\\hline\hline
$00$ & $[0,0,0,0,1]^T$ \\\hline
$01$ & $[0,0,0,1,0]^T$ \\\hline
$100$ & $[0,0,1,0,0]^T$ \\\hline
$101$ & $[0,1,0,0,0]^T$ \\\hline
$110$ & $[1,0,0,0,0]^T$ \\\hline
$111000$ & $[0,0,1,1,1]^T$ \\\hline
$111001$ & $[0,1,0,1,1]^T$ \\\hline
$111010$ & $[1,0,0,1,1]^T$ \\\hline
$111011$ & $[0,1,1,0,1]^T$ \\\hline
$111100$ & $[1,0,1,0,1]^T$ \\\hline
$111101$ & $[1,1,0,0,1]^T$ \\\hline
$1111100$ & $[0,1,1,1,0]^T$ \\\hline
$1111101$ & $[1,0,1,1,0]^T$ \\\hline
$1111110$ & $[1,1,0,1,0]^T$ \\\hline
$1111111$ & $[1,1,1,0,0]^T$ \\\hline
\end{tabular}
\end{center}
\end{table}
%%%%%%%%%%%%%%%%%%%%%%%%%%%%%%%%%%%%%%%%%%%%%%%%%%%%%%%%%%%%%%%%%

Due to the unequal length of bit strings, if an error occurs in the detector, the Huffman encoder at the receiver will map the incorrect symbol to a bit string that is different from the transmitted one and possibly of a different length. The mismatch in the bit length will result in an error propagation effect affecting all subsequent symbols regardless of whether they are correctly detected. Among other solutions, the following methods may help resolve this issue.
\begin{itemize}
\item[1)] {\it Frame-based operation:} In the frame (packet)-based operation of EE-HSSK scheme, the input bit sequence is divided into frames. Each frame is transmitted using multiple EE-HSSK modulation symbols. Different frames are separated by the all-zero (``idle") symbol which signals the beginning/end of the transmission of a frame. If the mismatch effect occurs, it affects the current frame but not the next frame.
\item[2)] {\it Channel coding and interleaving:} If the information sequence is protected by the channel encoder and the interleaver before entering the Huffman decoder, the bit errors resulted from the mismatch effect in a frame will be deinterleaved to non-contiguous locations and could be corrected by the channel decoder.
\end{itemize}

\begin{figure}[tb]
\begin{center}
\includegraphics[width=\columnwidth]{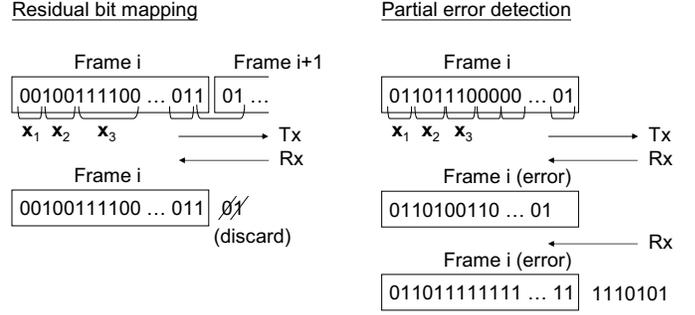}
\caption{Residual bit mapping and partial error detection in the frame-based EE-HSSK transmission.} \label{fig:HSSKmodel2}
\end{center}
\end{figure}

We consider the frame-based implementation of EE-HSSK enabled transmission in the uncoded scenarios. Operation details and an embedded capability of this implementation are described below.
\begin{itemize}
\item[1)] {\it Residual bit mapping:} In the frame-based implementation, if the residual bits at the end of a frame are not sufficient to map to an EE-HSSK symbol, the initial bits of the next frame are borrowed to complete the mapping. At the receiver, extra bits beyond a frame are discarded. This is illustrated in Fig.~\ref{fig:HSSKmodel2}, where the exemplary EE-HSSK mapping in Table \ref{tab:mod3} is adopted.
\item[2)] {\it Partial error detection (ED):} The frame-based implementation enables partial ED at no extra cost. Specifically, when the receiver recovers a bit sequence shorter than the frame size, or longer than the frame size plus the maximum length of EE-HSSK symbols minus one, the receiver detects an incorrect frame being received. This is illustrated in Fig.~\ref{fig:HSSKmodel2}. This embedded partial ED mechanism can be used in conjunction with automatic repeat request (ARQ) to improve the frame transmission performance, as will be demonstrated in Sec.~\ref{sec:simulation}.
\end{itemize}

%%%%%%%%%%%%%%%%%%%%%%%%%%%%%%%%%%%%%%%%%%%%%%%%%%%%%%%%%%%%%%%%%
%%%%%%%%%%%%%%%%%%%%%%%%%%%%%%%%%%%%%%%%%%%%%%%%%%%%%%%%%%%%%%%%%

\section{Simulation Results} \label{sec:simulation}

In this section, we show the simulation results for the performance of the proposed energy-efficient transmission scheme in comparison with the theoretical results and the conventional schemes. The frame size is set to 100 bits in simulation, and enough frames are simulated to count 300 frame errors. The lexicographic constellation design is adopted for GSSK modulation for computational efficiency without much loss of performance \cite{ChangLin11}. The detection algorithm is MAP for all schemes.

%%%%%%%%%%%%%%%%%%%%%%%%%%%%%%%%%%%%%%%%%%%%%%%%%%%%%%%%%%%%%%%%%
\begin{figure}[tb]
\begin{center}
\subfigure[]{
    \label{fig:power_entropy1}
    \includegraphics[width=\columnwidth]{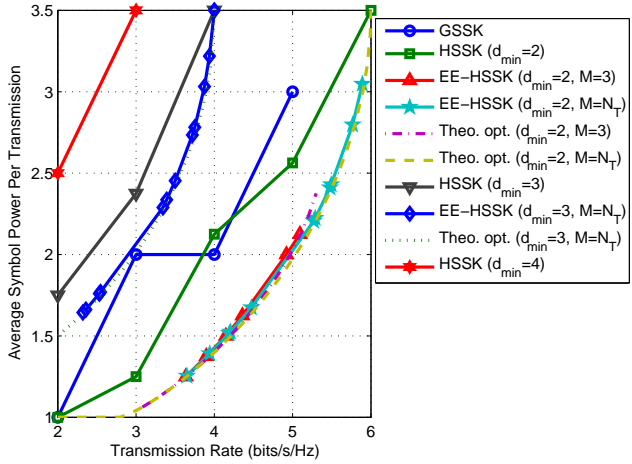}}
\subfigure[]{
    \label{fig:power_entropy2}
    \includegraphics[width=\columnwidth]{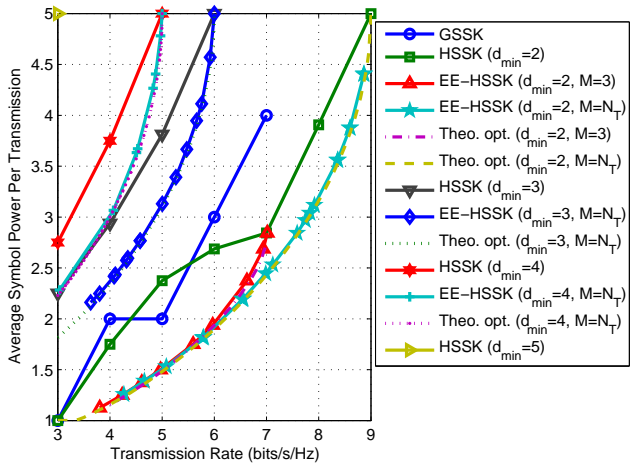}}
\caption{Average symbol power per transmission versus transmission rate for SSK-type modulation schemes in a system with (a) $N_T=7$ and (b) $N_T=10$.}
\label{fig:power_entropy}
\end{center}
\end{figure}
%%%%%%%%%%%%%%%%%%%%%%%%%%%%%%%%%%%%%%%%%%%%%%%%%%%%%%%%%%%%%%%%%

%%%%%%%%%%%%%%%%%%%%%%%%%%%%%%%%%%%%%%%%%%%%%%%%%%%%%%%%%%%%%%%%%
\begin{figure}[tb]
\begin{center}
\subfigure[]{
    \label{fig:ser_snr1}
    \includegraphics[width=\columnwidth]{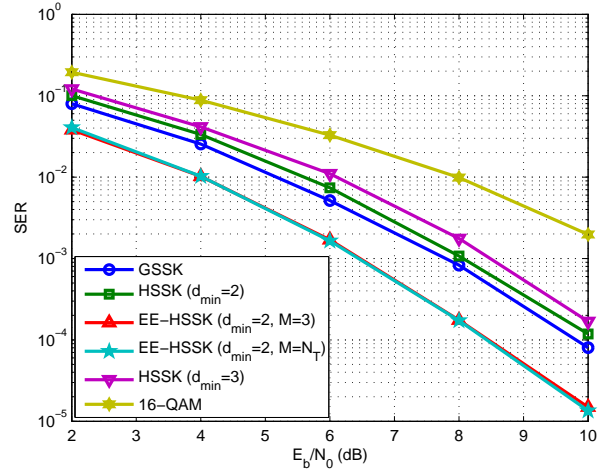}}
\subfigure[]{
    \label{fig:fer_snr1}
    \includegraphics[width=\columnwidth]{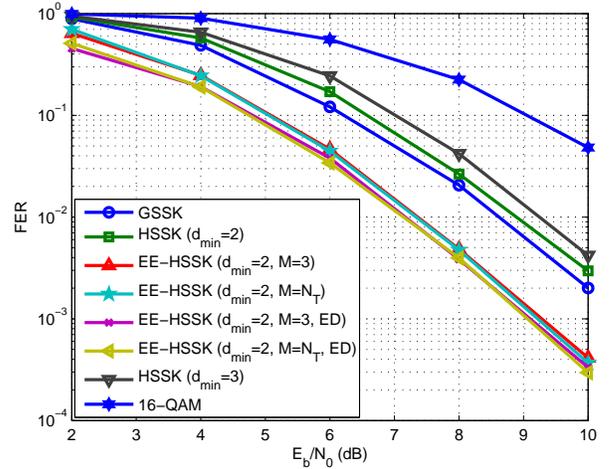}}
\caption{(a) SER and (b) FER performance for SSK-type modulation schemes ($7\times 7$ system) and 16-QAM ($1\times 7$ system), for 4 bits/s/Hz transmission.}
\label{fig:ser_fer_snr1}
\end{center}
\end{figure}
%%%%%%%%%%%%%%%%%%%%%%%%%%%%%%%%%%%%%%%%%%%%%%%%%%%%%%%%%%%%%%%%%

%%%%%%%%%%%%%%%%%%%%%%%%%%%%%%%%%%%%%%%%%%%%%%%%%%%%%%%%%%%%%%%%%
\begin{figure}[tb]
\begin{center}
\subfigure[]{
    \label{fig:ser_snr2}
    \includegraphics[width=\columnwidth]{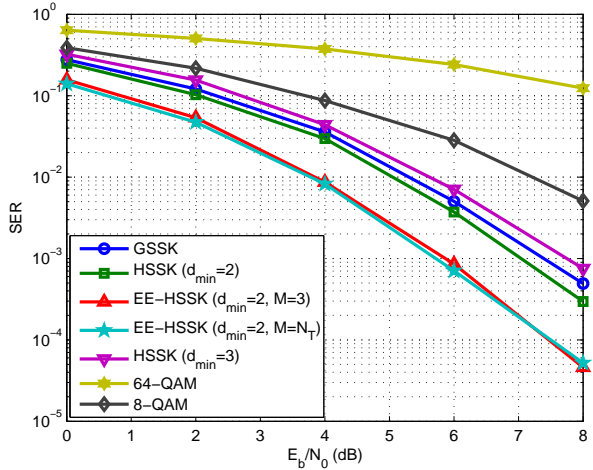}}
\subfigure[]{
    \label{fig:fer_snr2}
    \includegraphics[width=\columnwidth]{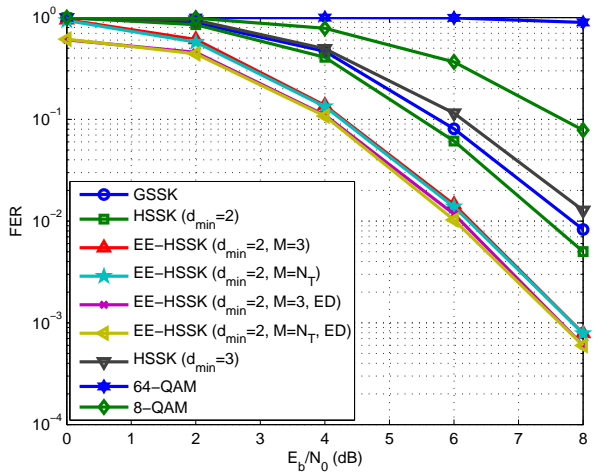}}
\caption{(a) SER and (b) FER performance for SSK-type modulation schemes ($10\times 10$ system), 64-QAM ($1\times 10$ system), and 8-QAM ($2\times 10$ system), for 6 bits/s/Hz transmission.}
\label{fig:ser_fer_snr2}
\end{center}
\end{figure}
%%%%%%%%%%%%%%%%%%%%%%%%%%%%%%%%%%%%%%%%%%%%%%%%%%%%%%%%%%%%%%%%%

In Fig.~\ref{fig:power_entropy}, we plot the average symbol power per transmission ${\mathbb E}[\tilde{{\mathbf x}}^H \tilde{{\mathbf x}}]$ versus the transmission rate for different SSK-type modulation schemes. First, we see that a more flexible utilization of the antenna indices allows HSSK to achieve different $d_{\mbox{\scriptsize min}}$ values while GSSK always has $d_{\mbox{\scriptsize min}}=2$. The adoption of a larger $d_{\mbox{\scriptsize min}}$ consumes higher average symbol power but contributes to a better error performance for HSSK (the tradeoff to be examined in Figs.~\ref{fig:ser_fer_snr1}--\ref{fig:ser_fer_snr2}). Second, HSSK can support transmission rates unachievable with GSSK, possibly using even lower average symbol power (e.g., 8 bits/s/Hz transmission for HSSK versus 7 bits/s/Hz transmission for GSSK in Fig.~\ref{fig:power_entropy2}) due to a more efficient utilization of the antenna indices. Third, the proposed EE-HSSK scheme maintains the advantages of HSSK modulation while demonstrates better energy efficiency as it nearly overlaps with the theoretical optimum for various $d_{\mbox{\scriptsize min}}$ configurations. The effect of imposing an RF-chain constraint $M<N_T$ is examined for the $d_{\mbox{\scriptsize min}}=2$ configuration. As can be seen, imposing a hardware constraint for EE-HSSK generally demonstrates negligible energy penalty. A bigger energy penalty is observed when the transmission rate approaches the capacity (maximum supportable rate) of the modulation scheme under the specified RF-chain constraint. For example, for a system with $N_T=10$ the capacity of EE-HSSK ($M=3$) is $130\big[-(1/130)\log_2(1/130)\big]=7.0224$ bits/s/Hz, incorporating all the ${10\choose 1}+{10\choose 3}=130$ symbols in ${\cal U}_1$ and ${\cal U}_3$ equiprobably in the alphabet. At the capacity-achieving rate of 7 bits/s/Hz, a minor energy gap is observed between adopting $M=3$ and $M=N_T=10$ for EE-HSSK, as shown in Fig.~\ref{fig:power_entropy2}. Also, at this rate EE-HSSK ($M=3$) and HSSK nearly overlap, because HSSK's alphabet with equiprobable symbols is nearly energy-optimal in this particular case. The simulation results motivate the use of EE-HSSK ($M<N_T$) scheme in practice to reduce the number of required RF chains from $N_T$ to $M$ at little loss of energy performance.

In Figs.~\ref{fig:ser_fer_snr1}--\ref{fig:ser_fer_snr2}, we plot the symbol-error-rate (SER) and frame-error-rate (FER) versus $E_b/N_0$ performance at fixed transmission rates, where $E_b$ represents the average energy of a single bit and is given by $E_s \cdot {\mathbb E}[\tilde{{\mathbf x}}^H \tilde{{\mathbf x}}]/m$ for $m$ bits/s/Hz transmission. The target transmission rates are 4 bits/s/Hz for a system with $N_T=7$ in Fig.~\ref{fig:ser_fer_snr1}, and 6 bits/s/Hz for a system with $N_T=10$ in Fig.~\ref{fig:ser_fer_snr2}. We consider symmetric systems ($N_T=N_R$) for SSK-type modulation schemes, although this is not a constraint on the proposed energy-efficient transmission strategy. We see that EE-HSSK with $M=3$ and $M=N_T$ have almost identical performance, both outperforming the original GSSK/HSSK and the conventional QAM (with gray-coded bit mapping). HSSK with $d_{\mbox{\scriptsize min}}=3$ configuration exhibits a slight disadvantage of performance as compared with $d_{\mbox{\scriptsize min}}=2$ configuration, suggesting that in these particular scenarios the improved error performance due to a larger $d_{\mbox{\scriptsize min}}$ does not outweigh the extra power spent for creating a larger $d_{\mbox{\scriptsize min}}$. The performance gap between EE-HSSK and GSSK/HSSK is about 1.5--2 dB at $\mbox{SER}=10^{-3}$ and $\mbox{FER}=10^{-2}$ in both Figs.~\ref{fig:ser_fer_snr1} and \ref{fig:ser_fer_snr2}. The embedded ED mechanism working in conjunction with ARQ-based transmission can further improve the FER performance of EE-HSSK. With the implementation of ED, a frame in error, when detected, is retransmitted so that with high probability it will be received correctly with joint detection of the two transmissions (we assume that the first retransmission is always successful in our simulation). The performance improvement is however slight, since not all frame error patterns can be detected by the embedded ED mechanism (e.g., if no bit-length mismatch occurs, the frame error cannot be detected).

\section{Conclusion} \label{sec:conclusion}

The energy efficiency of SSK-type modulation enabled transmission over MIMO wireless channels has been studied. A theoretical framework for the design of energy-efficient communication forms the foundation for the development of a practical energy-efficient modulation scheme. The proposed modulation scheme incorporates a novel use of the Hamming code and Huffman code techniques in the alphabet and bit-mapping designs. The proposed modulation scheme carries an energy-saving potential unachievable in previous schemes, as demonstrated by extensive simulations, and shows great promise for next-generation green MIMO communications.

\appendix[Properties of GSSK Modulation]

%%%%%%%%%%%%%%%%%%%%%%%%%%%%%%%%%%%%%%%%%%%%%%%%%%%%%%%%%%%%%%%%%
\begin{figure}[tb]
\begin{center}
\includegraphics[width=0.8\columnwidth]{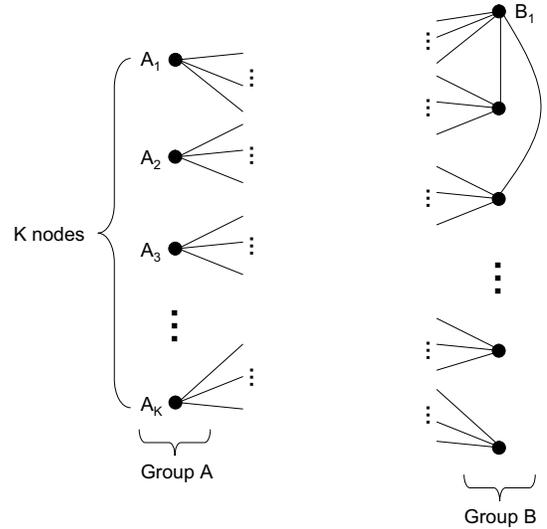}
\caption{Graph constructed for Proof of Theorem 1.} \label{fig:GSSKgraph}
\end{center}
\end{figure}
%%%%%%%%%%%%%%%%%%%%%%%%%%%%%%%%%%%%%%%%%%%%%%%%%%%%%%%%%%%%%%%%%

Recall that, to support $m$ bits/s/Hz transmission in a system with $N_T$ transmit antennas, GSSK modulation adopts a modulation alphabet ${\cal A}^{(\mbox{\scriptsize GSSK})}$ with $2^m$ $N_T\times 1$ symbols each comprised of $n_t$ 1's and $N_T-n_t$ 0's, where $n_t$ is chosen such that
\begin{equation} \label{eq:GSSK}
{N_T \choose n_t-1} <  \left|{\cal A}^{(\mbox{\scriptsize GSSK})}\right|=2^m \leq {N_T \choose n_t} \quad\mbox{and}\quad n_t\leq \frac{N_T}{2}.
\end{equation}

{\it Lemma 1:} The Hamming distance between arbitrary two distinct GSSK symbols ${\mathbf x}_i, {\mathbf x}_j\in{\cal A}^{(\mbox{\scriptsize GSSK})}$ is a positive even number, i.e., $d({\mathbf x}_i, {\mathbf x}_j)=2,4,6,\ldots$.

\begin{IEEEproof}
Assume ${\mathbf x}_i$ and ${\mathbf x}_j$ differ in $k$ positions, $1\leq k\leq N_T$. Thus, $d({\mathbf x}_i, {\mathbf x}_j)=k$. Since the numbers of 1's in ${\mathbf x}_i$ and ${\mathbf x}_j$ are equal, and ${\mathbf x}_i$ and ${\mathbf x}_j$ are identical in the other $N_T-k$ positions, ${\mathbf x}_i$ and ${\mathbf x}_j$ contain the same number of 1's, say $l$, where $1\leq l\leq n_t$, in these $k$ positions. Since ${\mathbf x}_i$ and ${\mathbf x}_j$ differ in these $k$ positions, we must have $k=2l$.
\end{IEEEproof}

{\it Theorem 1:} For any ${\cal A}^{(\mbox{\scriptsize GSSK})}$, the minimum Hamming distance between arbitrary two distinct GSSK symbols is 2, i.e., $d_{\mbox{\scriptsize min}}=\min_{{\mathbf x}_i\neq {\mathbf x}_j\in{\cal A}^{(\mbox{\scriptsize GSSK})}}d({\mathbf x}_i, {\mathbf x}_j)=2$.

\begin{IEEEproof}
If $n_t=1$, it is straightforward that GSSK (i.e., SSK) modulation has $d_{\mbox{\scriptsize min}}=2$. Therefore, we consider $n_t\geq 2$. Assume the opposite; that is, by Lemma 1, GSSK modulation has $d_{\mbox{\scriptsize min}}=4$. Let ${\cal U}_{n_t}$ be the set of symbol candidates from which the GSSK modulation symbols are selected, where $|{\cal U}_{n_t}|={N_T \choose n_t}$. We construct an undirected graph with ${N_T\choose n_t}$ nodes, each node representing a symbol candidate in ${\cal U}_{n_t}$. Two nodes are connected by an edge if they represent symbols that have a Hamming distance of 2, and unconnected otherwise. Nodes in the graph are divided into two groups, where group $A$ contains $K\triangleq 2^m$ nodes that represent the GSSK modulation symbols and group $B$ contains the rest of nodes, as shown in Fig.~\ref{fig:GSSKgraph}. Since we assume that GSSK modulation has $d_{\mbox{\scriptsize min}}=4$, there is no edge between nodes in group $A$. There is, however, possibly an edge between a node in group $A$ and a node in group $B$, as well as between two nodes in group $B$, as depicted in Fig.~\ref{fig:GSSKgraph}. Suppose that, without loss of generality, node $A_1$ in group $A$ represents the following symbol\footnote{For notational convenience, we use $101$ to denote symbol $[1,0,1]^T$, etc.}
\begin{equation} \label{eq:GSSK2}
\overbrace{1 1 1 \cdots 1}^{n_t-1} \overbrace{0 0 0 \cdots 0}^{N_T-n_t} 1.
\end{equation}
Symbols in ${\cal U}_{n_t}$ that have a Hamming distance of 2 from the symbol in (\ref{eq:GSSK2}) must be represented by nodes in group $B$. These symbols can be produced by switching positions of an 1 and a 0 in (\ref{eq:GSSK2}), and therefore there are $n_t(N_T-n_t)$ such symbols. By our graph construction rule, there is an edge between node $A_1$ and each of these $n_t(N_T-n_t)$ nodes in group $B$. Thus, node $A_1$ has $n_t(N_T-n_t)$ edges going out to group $B$. Likewise, for each node $A_2, A_3, \ldots, A_K$ in group $A$, there are also $n_t(N_T-n_t)$ edges going out to group $B$. As a result, the total number of edges going out from group $A$ to group $B$ is $K n_t(N_T-n_t)$.

Suppose that, without loss of generality, node $B_1$ in group $B$ represents the following symbol
\begin{equation} \label{eq:GSSK3}
\overbrace{1 1 1 \cdots 1}^{n_t} \overbrace{0 0 0 \cdots 0}^{N_T-n_t}.
\end{equation}
If a node in group $A$ and node $B_1$ are connected by an edge, this node in group $A$ represents a symbol that has a Hamming distance of 2 from the symbol in (\ref{eq:GSSK3}) and can be produced by switching positions of an 1 and a 0 in (\ref{eq:GSSK3}). That is, it is a concatenation of a row in ${\mathbf G}_1$ and a row in ${\mathbf G}_2$:
\begin{equation} \label{eq:GSSK4}
\underbrace{
\begin{bmatrix}
111\cdots 110 \\
111\cdots 101 \\
111\cdots 011 \\
\vdots \\
011\cdots 111
\end{bmatrix}
}_{{\mathbf G}_1: n_t\times n_t}
\underbrace{
\begin{bmatrix}
000\cdots 001 \\
000\cdots 010 \\
000\cdots 100 \\
\vdots \\
100\cdots 000
\end{bmatrix}
}_{{\mathbf G}_2: (N_T-n_t)\times (N_T-n_t)}.
\end{equation}
Suppose that there are two nodes in group $A$ that are each connected with node $B_1$ by an edge. Then, these two nodes must each represent a symbol comprised of a row in ${\mathbf G}_1$ and a row in ${\mathbf G}_2$, as discussed above. Furthermore, since these two nodes in group $A$ represent symbols with a Hamming distance of 4 or more in between, they must each represent a symbol comprised of a different row in ${\mathbf G}_1$ {\it and} a different row in ${\mathbf G}_2$ because otherwise they would have a Hamming distance of 2 rather than 4. For this reason, node $B_1$ can be connected by an edge with at most $\min(n_t, N_T-n_t)=n_t$ nodes in group $A$, or equivalently, node $B_1$ has at most $n_t$ edges going out to group $A$. Likewise, each of the other nodes in group $B$ has at most $n_t$ edges going out to group $A$.

Since the total number of edges going out from group $A$ to group $B$ is $K n_t(N_T-n_t)$, and each node in group $B$ can have at most $n_t$ edges going out to group $A$, there must be at least $K n_t(N_T-n_t)/n_t=K(N_T-n_t)$ nodes in group $B$. Since the combined total number of nodes in group $A$ and group $B$ is ${N_T\choose n_t}$, we have
\begin{equation}
\underbrace{K}_{\mbox{\# nodes in group $A$}}+\underbrace{K(N_T-n_t)}_{\mbox{min. \# nodes in group $B$}}\leq {N_T\choose n_t}
\end{equation}
or
\begin{eqnarray}
K &\leq& \frac{1}{N_T-n_t+1}{N_T\choose n_t} \nonumber\\
&=& \frac{1}{n_t}{N_T\choose n_t-1}. \label{eq:GSSK5}
\end{eqnarray}
Since $n_t\geq 2$, (\ref{eq:GSSK5}) and (\ref{eq:GSSK}) are in contradiction, which completes the proof.
\end{IEEEproof}

{\it Corollary 1:} For any ${\cal A}^{(\mbox{\scriptsize GSSK})}$, the maximum number of GSSK symbols with a pairwise Hamming distance of 4 or more is $\left\lfloor 1/(N_T-n_t+1){N_T\choose n_t}\right\rfloor$.

\begin{IEEEproof}
This result follows from (\ref{eq:GSSK5}) in Proof of Theorem 1.
\end{IEEEproof}

{\it Corollary 2:} For any ${\cal A}^{(\mbox{\scriptsize GSSK})}$, the maximum number of GSSK symbols with a pairwise Hamming distance of 6 or more is $\left\lfloor 1/(N_T n_t-n_t^2+1){N_T\choose n_t}\right\rfloor$.

\begin{IEEEproof}
Following Proof of Theorem 1 and considering that the $K$ nodes in group $A$ have a pairwise Hamming distance of at least 6, it is clear that node $B_1$ can be connected by an edge with at most one node in group $A$. This is because otherwise the two or more nodes in group $A$ connected with $B_1$ must each represent a symbol comprised of a row in ${\mathbf G}_1$ and a row in ${\mathbf G}_2$ in (\ref{eq:GSSK4}) and therefore cannot have a Hamming distance of 6 or more in between. Since each node in group $B$ has at most one edge going out to group $A$, there must be at least $K n_t(N_T-n_t)$ nodes in group $B$. Thus,
\begin{equation}
K + K n_t(N_T-n_t)\leq {N_T\choose n_t}
\end{equation}
or equivalently
\begin{equation}
K\leq \frac{1}{N_T n_t-n_t^2+1} {N_T\choose n_t}.
\end{equation}
\end{IEEEproof}

Corollary 1 and Corollary 2 have an implication for GSSK-based transmission in large MIMO systems. Specifically, for any ${\cal A}^{(\mbox{\scriptsize GSSK})}$ the number of GSSK symbols with a pairwise Hamming distance of 4 (or 6) or more compared with the total number of symbol candidates $|{\cal U}_{n_t}|={N_T \choose n_t}$ is no more than the ratio of $2/N_T$ (or $4/N_T^2$) when $N_T$ is large and $n_t\approx N_T/2$. This suggests that an increasingly small subset of the available symbol candidates can have a pairwise Hamming distance larger than 2 as $N_T$ increases, a result of the inefficiency of the fixed $n_t$ construction in GSSK modulation.

\bibliographystyle{IEEEtran}  % styles: plain, abbrv, alpha, astron ...
\bibliography{IEEEabrv,mybib}

\begin{thebibliography}{10}
\providecommand{\url}[1]{#1}
\csname url@rmstyle\endcsname
\providecommand{\newblock}{\relax}
\providecommand{\bibinfo}[2]{#2}
\providecommand\BIBentrySTDinterwordspacing{\spaceskip=0pt\relax}
\providecommand\BIBentryALTinterwordstretchfactor{4}
\providecommand\BIBentryALTinterwordspacing{\spaceskip=\fontdimen2\font plus
\BIBentryALTinterwordstretchfactor\fontdimen3\font minus
  \fontdimen4\font\relax}
\providecommand\BIBforeignlanguage[2]{{%
\expandafter\ifx\csname l@#1\endcsname\relax
\typeout{** WARNING: IEEEtran.bst: No hyphenation pattern has been}%
\typeout{** loaded for the language `#1'. Using the pattern for}%
\typeout{** the default language instead.}%
\else
\language=\csname l@#1\endcsname
\fi
#2}}

\bibitem{MeslehHaas08}
R.~Y. Mesleh, H.~Haas, S.~{Sinanovi\'{c}}, C.~W. Ahn, and S.~Yun, ``Spatial
  modulation,'' \emph{{IEEE} Trans. Veh. Technol.}, vol.~57, no.~4, pp.
  2228--2241, July 2008.

\bibitem{YounisSerafimovski10}
A.~Younis, N.~Serafimovski, R.~Mesleh, and H.~Haas, ``Generalised spatial
  modulation,'' in \emph{2010 Conference Record of the Forty Fourth Asilomar
  Conference on Signals, Systems and Computers (ASILOMAR)}, Nov. 2010, pp.
  1498--1502.

\bibitem{QuWang11}
Q.~Qu, A.~Wang, Z.~Nie, and J.~Zheng, ``Block mapping spatial modulation scheme
  for {MIMO} systems,'' \emph{The Journal of China Universities of Posts and
  Telecommunications}, vol.~18, no.~5, pp. 30--36, Oct. 2011.

\bibitem{WangJia12}
J.~Wang, S.~Jia, and J.~Song, ``Generalised spatial modulation system with
  multiple active transmit antennas and low complexity detection scheme,''
  \emph{{IEEE} Trans. Wireless Commun.}, vol.~11, no.~4, pp. 1605--1615, Apr.
  2012.

\bibitem{FuHou10}
J.~Fu, C.~Hou, W.~Xiang, L.~Yan, and Y.~Hou, ``Generalised spatial modulation
  with multiple active transmit antennas,'' in \emph{Proc. IEEE Globecom
  Workshops}, Dec. 2010, pp. 839--844.

\bibitem{SerafimovskiDiRenzo10}
N.~Serafimovski, M.~{Di Renzo}, S.~Sinanovic, R.~Y. Mesleh, and H.~Haas,
  ``Fractional bit encoded spatial modulation {(FBE-SM)},'' \emph{{IEEE}
  Commun. Lett.}, vol.~14, no.~5, pp. 429--431, May 2010.

\bibitem{YangAissa11}
Y.~Yang and S.~Aissa, ``Bit-padding information guided channel hopping,''
  \emph{{IEEE} Commun. Lett.}, vol.~15, no.~2, pp. 163--165, Feb. 2011.

\bibitem{MeslehDiRenzo10}
R.~Mesleh, M.~{Di Renzo}, H.~Haas, and P.~M. Grant, ``Trellis coded spatial
  modulation,'' \emph{{IEEE} Trans. Wireless Commun.}, vol.~9, no.~7, pp.
  2349--2361, July 2010.

\bibitem{BasarAygolu11}
E.~Basar, U.~Aygolu, E.~Panayirci, and H.~V. Poor, ``New trellis code design
  for spatial modulation,'' \emph{{IEEE} Trans. Wireless Commun.}, vol.~10,
  no.~8, pp. 2670--2680, Aug. 2011.

\bibitem{BasarAygolu11_2}
------, ``Space-time block coded spatial modulation,'' \emph{{IEEE} Trans.
  Commun.}, vol.~59, no.~3, pp. 823--832, Mar. 2011.

\bibitem{HandteMuller09}
T.~Handte, A.~{M\"{u}ller}, and J.~Speidel, ``{BER} analysis and optimization
  of generalized spatial modulation in correlated fading channels,'' in
  \emph{Proc. IEEE VTC 2009-Fall}, Sept. 2009, pp. 1--5.

\bibitem{DiRenzoHaas10}
M.~{Di Renzo} and H.~Haas, ``Performance analysis of spatial modulation,'' in
  \emph{Proc. IEEE Chinacom}, Aug. 2010, pp. 1--7.

\bibitem{DiRenzoHaas10_2}
------, ``A general framework for performance analysis of space shift keying
  {(SSK)} modulation for {MISO} correlated {Nakagami-$m$} fading channels,''
  \emph{{IEEE} Trans. Commun.}, vol.~58, no.~9, pp. 2590--2603, Sept. 2010.

\bibitem{DiRenzoHaas11_2}
------, ``Space shift keying {(SSK---)} {MIMO} over correlated {Rician} fading
  channels: Performance analysis and a new method for transmit-diversity,''
  \emph{{IEEE} Trans. Commun.}, vol.~59, no.~1, pp. 116--129, Sept. 2011.

\bibitem{DiRenzoHaas10_3}
------, ``Performance comparison of different spatial modulation schemes in
  correlated fading channels,'' in \emph{Proc. IEEE ICC}, May 2010, pp. 1--6.

\bibitem{JeganathanGhrayeb09}
J.~Jeganathan, A.~Ghrayeb, L.~Szczecinski, and A.~Ceron, ``Space shift keying
  modulation for {MIMO} channels,'' \emph{{IEEE} Trans. Wireless Commun.},
  vol.~8, no.~7, pp. 3692--3703, July 2009.

\bibitem{JeganathanGhrayeb08}
J.~Jeganathan, A.~Ghrayeb, and L.~Szczecinski, ``Generalized space shift keying
  modulation for {MIMO} channels,'' in \emph{Proc. IEEE PIMRC}, Sept. 2008, pp.
  1--5.

\bibitem{ChangLin11}
R.~Y. Chang, S.-J. Lin, and W.-H. Chung, ``New space shift keying modulation
  with {Hamming} code-aided constellation design,'' \emph{{IEEE} Wireless
  Commun. Lett.}, vol.~1, no.~1, pp. 2--5, Feb. 2012.

\bibitem{LinCostello04}
S.~Lin and D.~J. Costello, \emph{Error control coding}, 2nd~ed.\hskip 1em plus
  0.5em minus 0.4em\relax Prentice Hall, 2004.

\bibitem{SugiuraChen10}
S.~Sugiura, S.~Chen, and L.~Hanzo, ``Coherent and differential space-time shift
  keying: A dispersion matrix approach,'' \emph{{IEEE} Trans. Commun.},
  vol.~58, no.~11, pp. 3219--3230, Nov. 2010.

\bibitem{SugiuraChen11}
------, ``Generalized space-time shift keying designed for flexible diversity-,
  multiplexing- and complexity-tradeoffs,'' \emph{{IEEE} Trans. Wireless
  Commun.}, vol.~10, no.~4, pp. 1144--1153, Apr. 2011.

\bibitem{YangXu11}
D.~Yang, C.~Xu, L.-L. Yang, and L.~Hanzo, ``Transmit-diversity-assisted
  space-shift keying for colocated and distributed/cooperative {MIMO}
  elements,'' \emph{{IEEE} Trans. Veh. Technol.}, vol.~60, no.~6, pp.
  2864--2869, July 2011.

\bibitem{DiRenzoHaas10_4}
M.~{Di Renzo} and H.~Haas, ``Improving the performance of space shift keying
  {(SSK)} modulation via opportunistic power allocation,'' \emph{{IEEE} Commun.
  Lett.}, vol.~14, no.~6, pp. 500--502, June 2010.

\bibitem{MohammadiGhannouchi11}
A.~Mohammadi and F.~M. Ghannouchi, ``Single {RF} front-end {MIMO}
  transceivers,'' \emph{{IEEE} Commun. Mag.}, vol.~49, no.~12, pp. 104--109,
  Dec. 2011.

\bibitem{SanayeiNosratinia04}
S.~Sanayei and A.~Nosratinia, ``Antenna selection in {MIMO} systems,''
  \emph{{IEEE} Commun. Mag.}, vol.~42, no.~10, pp. 68--73, Oct. 2004.

\bibitem{DiRenzoHaas11}
M.~{Di Renzo}, H.~Haas, and P.~M. Grant, ``Spatial modulation for
  multiple-antenna wireless systems: A survey,'' \emph{{IEEE} Commun. Mag.},
  vol.~49, no.~12, pp. 182--191, Dec. 2011.

\bibitem{JeganathanGhrayeb08_2}
J.~Jeganathan, A.~Ghrayeb, and L.~Szczecinski, ``Spatial modulation: Optimal
  detection and performance analysis,'' \emph{{IEEE} Commun. Lett.}, vol.~12,
  no.~8, pp. 545--547, Aug. 2008.

\bibitem{Bertsekas99}
D.~P. Bertsekas, \emph{Nonlinear Programming}, 2nd~ed.\hskip 1em plus 0.5em
  minus 0.4em\relax Athena Scientific, 1999.

\bibitem{Huffman52}
D.~A. Huffman, ``A method for the construction of minimum-redundancy codes,''
  \emph{Proceedings of the IRE}, vol.~40, no.~9, pp. 1098--1101, Sept. 1952.

\end{thebibliography}

\begin{IEEEbiography}[{\includegraphics[width=1in,height=1.25in,clip,keepaspectratio]{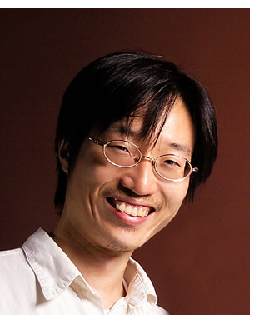}}]
{Ronald Y. Chang} received the B.S. degree in electrical engineering from the National Tsing Hua University, Hsinchu, Taiwan, in 2000, the M.S. degree in electronics engineering from the National Chiao Tung University, Hsinchu, in 2002, and the Ph.D. degree in electrical engineering from the University of Southern California (USC), Los Angeles, in 2008. 

From 2002 to 2003, he was with the Industrial Technology Research Institute, Hsinchu. Since 2004, he has conducted research with USC as well as with the Mitsubishi Electric Research Laboratories, Cambridge, MA. Since 2010, he has been with the Research Center for Information Technology Innovation at Academia Sinica, Taipei, Taiwan. His research interests include resource allocation, interference management, detection and estimation, cognitive radio and cooperative communications for wireless communications and networking. He is a recipient of the Best Paper Award from IEEE WCNC 2012, and the holder of four awarded and one pending U.S. patents.
\end{IEEEbiography}

\vspace*{-1.5\baselineskip}

\begin{IEEEbiography}[{\includegraphics[width=1in,height=1.25in,clip,keepaspectratio]{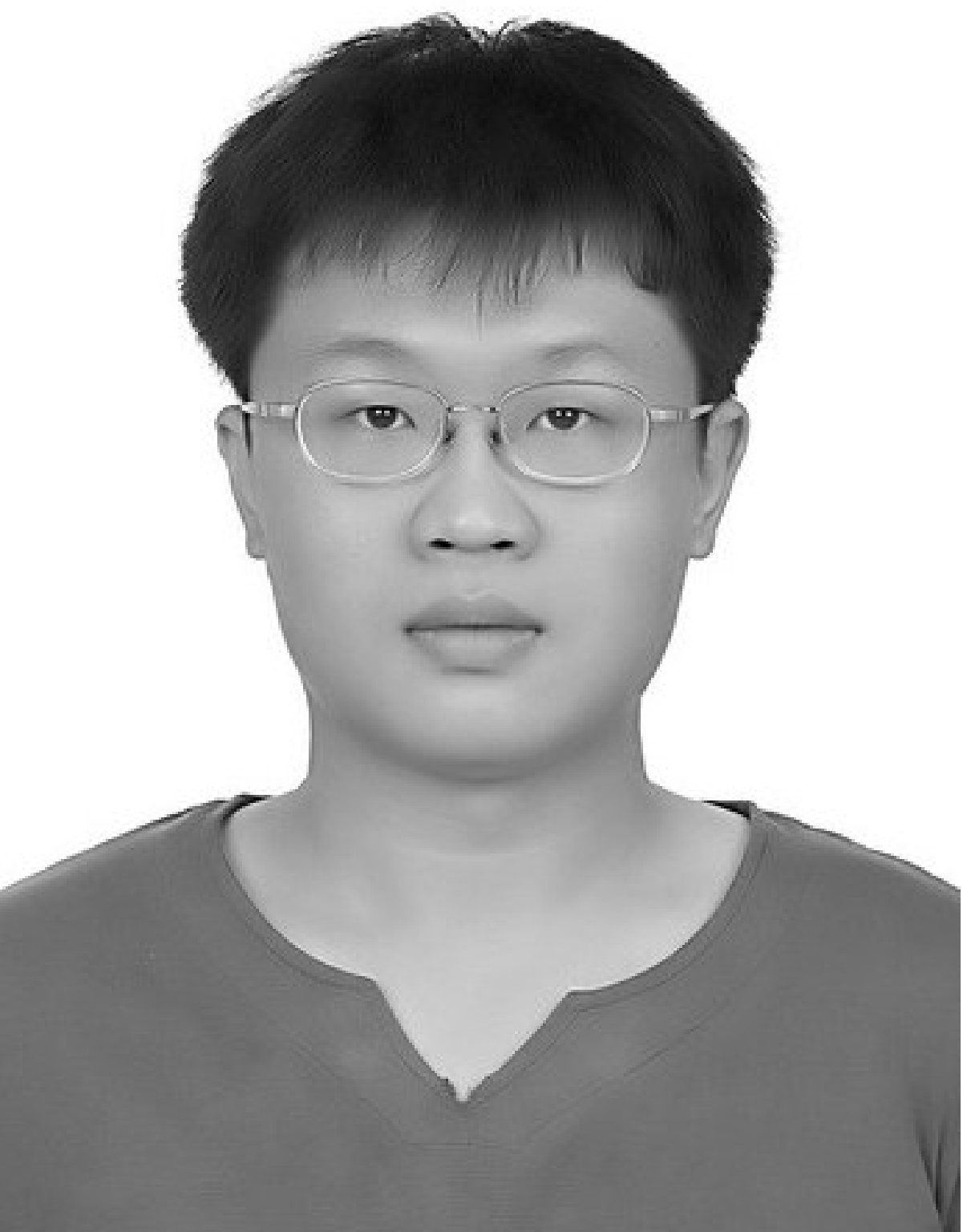}}]
{Sian-Jheng Lin} was born in Taiwan. He received the B.S., M.S., and Ph.D. degrees in computer science from National Chiao Tung University, in 2004, 2006, and 2010, respectively. He is currently a postdoctoral fellow with the Research Center for Information Technology Innovation, Academia Sinica. His recent research interests include data hiding, error control coding, modulation, and secret sharing.
\end{IEEEbiography}

\vspace*{-1.5\baselineskip}

\begin{IEEEbiography}[{\includegraphics[width=1in,height=1.25in,clip,keepaspectratio]{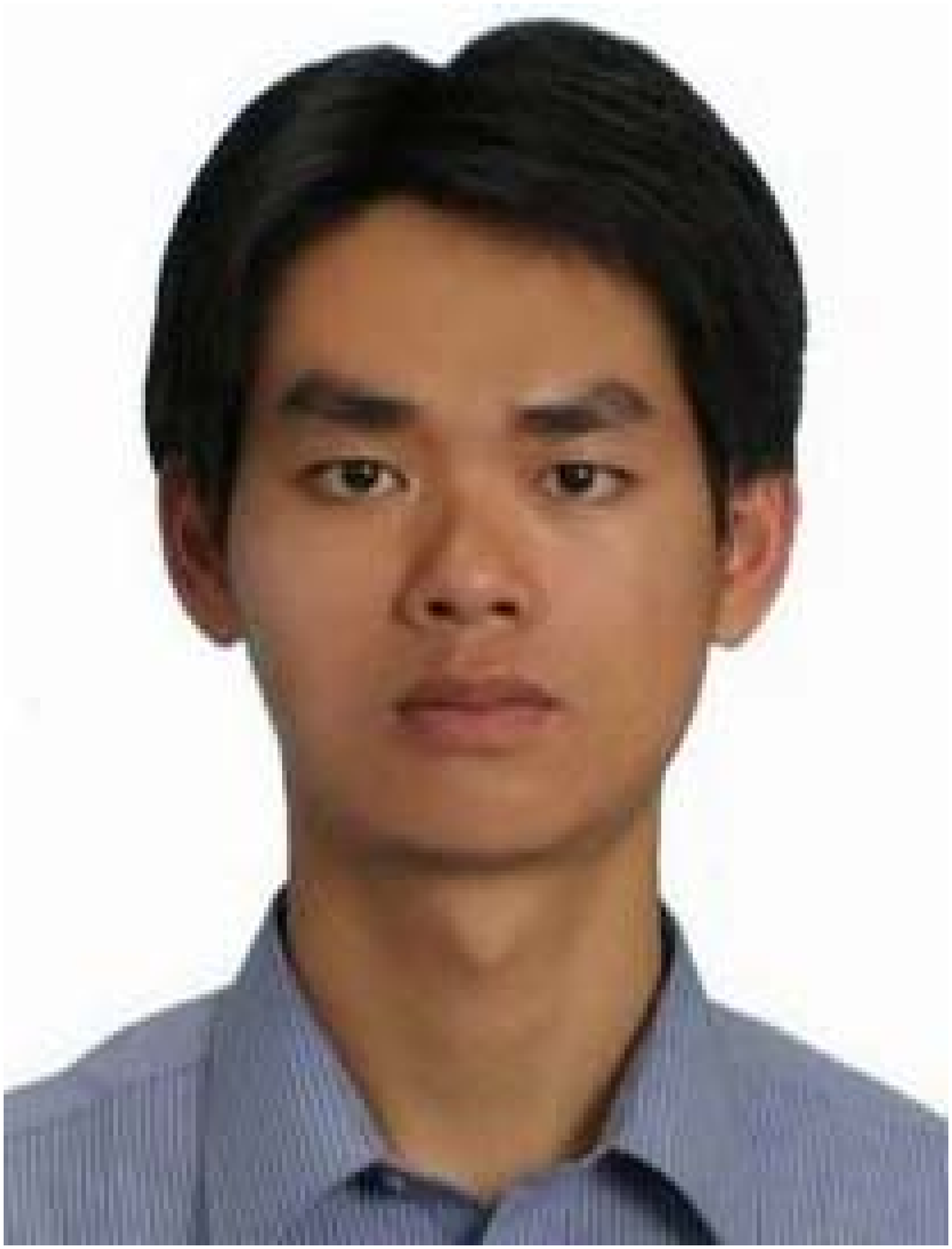}}]
{Wei-Ho Chung} (M'11) was born in Kaohsiung, Taiwan, in 1978. He received the B.Sc. and M.Sc. degrees in Electrical Engineering from National Taiwan University, Taipei City, Taiwan, in 2000 and 2002, respectively. From 2005 to 2009, he was with the Electrical Engineering Department at University of California, Los Angeles, where he obtained his Ph.D. degree. From 2000 to 2002, he worked on routing protocols in the mobile ad hoc networks in the M.Sc. program in National Taiwan University. From 2002 to 2005, he was a system engineer at ChungHwa Telecommunications Company, where he worked on data networks. In 2008, he was an research intern working on CDMA systems in Qualcomm, Inc. His research interests include communications, signal processing, and networks. Dr. Chung received the Taiwan Merit Scholarship from 2005 to 2009, and the Best Paper Award in IEEE WCNC 2012. Dr. Chung has been an assistant research fellow in the Research Center for Information Technology Innovation at Academia Sinica, Taiwan, since January 2010.
\end{IEEEbiography}

\end{document}